# Reversible phase transition in laser-shocked 3Y-TZP ceramics observed via nanosecond time-resolved X-ray diffraction

Jianbo Hu,[1,5,a)] Kouhei Ichiyanagi,[2] Hiroshi Takahashi,[1] Hiroaki Koguchi,[1] Takeaki Akasaka,[1] Nobuaki Kawai,[3] Shunsuke Nozawa,[4] Tokushi Sato,[4] Yuji C. Sasaki,[2] Shin-ichi Adachi,[4] and Kazutaka G. Nakamura[1,b)]

[1] *Materials and Structures Laboratory, Tokyo Institute of Technology, R3-10, 4259 Nagatsuta, Yokohama 226-8503, Japan*

[2] *Graduate School of Frontier Sciences, the University of Tokyo, 5-1-5 Kashiwanoha, Kashiwa, Chiba 277-8562, Japan*

[3] *Institute of Space and Astronautical Science, Japan Aerospace Exploration Agency, Yoshinodai 3-1-1, Sagamihara 229-8510, Japan*

[4] *Photon Factory, High Energy Accelerator Research Organization, 1-1 Oho, Tsukuba, Ibaraki 305-0801, Japan*

[5] *Laboratory for Shock Wave and Detonation Physics Research, Institute of Fluid Physics, Chinese Academy of Engineering Physics, P. O. Box 919-102 Mianyang, Sichuan 621900, China*

The high-pressure phase stability of the metastable tetragonal zirconia is still under debate. The transition dynamics of shocked $Y_2O_3$ (3 mol%) stabilized tetragonal zirconia ceramics under laser-shock compression has been directly studied using nanosecond time-resolved X-ray diffraction. The martensitic phase transformation to the monoclinic phase, which is the stable phase for pure zirconia at ambient pressure and room temperature, has been observed during compression at 5 GPa within 20 ns without any intermediates. This monoclinic phase reverts back to the tetragonal phase during pressure release. The results imply that the stabilization effect due to the addition of $Y_2O_3$ is to some extent negated by the shear stress under compression.

Investigation on phase transitions of materials under shock compression has bottlenecked due to the limitation of traditional experimental approaches which provide incomplete information on structural dynamics. Shock-recovery experiments, for example, only obtain end-state microscopic information used to infer dynamic behaviors at post-shocked states.[1-3] *In situ* bulk property measurements, such as velocity interferometer and stress gauge, in real-time detect the exact post-shocked state but cannot determine its structure without the help from static high-pressure experiments or theoretical computations.[4-6] Furthermore, traditional techniques have limited applications in studying phase transitions which are irreversible or associated with negligible volume change. In the latter situation, usually no step feature used to identify phase transitions can be observed in Hugoniot curves or particle velocity profiles. The progress of time-resolved X-ray diffraction technique makes us possible to monitor the on-the-fly state in the atomic scale.[7-11] Utilizing this ability to investigate dynamic behaviors of materials under shock compression provided considerable insight into microscopic mechanism of phase transformation, which is inaccessible by traditional approaches. In this report, we present our time-resolved X-ray diffraction observation on structural dynamics of laser-shocked $Y_2O_3$ (3 mol%) stabilized tetragonal zirconia polycrystalline (3Y-TZP) ceramics.

3Y-TZP ceramics, consisting of almost 100 % tetragonal phase, has been extensively studied due to its excellent mechanical properties, such as strength and toughness.[12] In contrast to the well-recognized phase diagram of pure zirconia,[13-15] the phase stability of 3Y-TZP under high pressure is still questionable, although considerable efforts have been devoted to this topic.[16-23] Static and dynamic high pressure experiments provided appreciably different results. In the former the $Y_2O_3$ stabilized tetragonal zirconia under high pressure transformed into a disorder structure [16] or an orthorhombic II phase via the monoclinic phase.[17, 18] While in the latter, including both Hugoniot measurements[19-21] and shock-recovery experiments,[22, 23] it was inferred that the tetragonal zirconia directly transformed into the orthorhombic II phase during compression or transformed into a quenchable monoclinic phase during releasing from a shocked state. As mentioned above, shock-wave experiments with traditional techniques only provide indirect information of the post-shocked state. Therefore, it is worth paying more attention on structural dynamics of 3Y-TZP ceramics under shock-compression. On the other hand, the possible transition from tetragonal to monoclinic, associated with transformation toughening, is of importance in various engineering application.[24] The transformation path of the tetragonal-monoclinic transition, however, is not well established. Several possible intermediates have been proposed to account for such transformation.[25-27] Due to the speed of this transformation it was experimentally difficult to resolve the transient structures in the past. Ultrafast time-resolved X-ray diffraction technique provides possibility to solve this problem.

The experiment was performed using the beamline NW14A at the Photon Factory Advanced Ring, KEK.[28, 29] A single-shot laser pump-X-ray probe scheme was applied to capture the structural change of laser-shocked 3Y-TZP ceramics as a function of the pump-probe delay Δt controlled by a delay generator, schematically shown in Fig. 1(a). The X-ray pulse, from an undulator with the period length of 20 mm (U20), has the pulse duration of about 100 ps. The laser pulse from a Q-switched Nd: YAG laser (Continuum, Powerlite 8000) was used for shock compression, with the wavelength, pulse width, and pulse energy of 1064 nm, 10 ns, and 700 mJ, respectively. The synchronization of laser and X-ray pulses using delay generator (DG535, Stanford Research System, Inc.) is realized with a time jitter less than 2 ns.[28] The peak energy, bandwidth and flux of the X-ray probe pulse were 15.6 KeV, 4.4 % and around $3\times10^8$ photons/pulse, respectively.[29] The target assembly, as shown in Fig. 1(b), was composed of three layers: the backup plastic film (25 μm), the Al ablation film (1 μm) and the 3Y-TZP plate (50 μm, Tosoh Co.). Such geometry confined laser-generated plasma at the interface of Al and plastic films, driving the pressure pulse through the Al film into the sample. The X-ray probe was normally incident upon the target assembly, focusing on a spot of 0.49×0.24 mm², and the laser pump was slightly deviated from the probe in angle with a focal spot of 0.4×0.4 mm². The Debye-Scherrer diffraction patterns were recorded on an integrating charge-coupled device detector (MarCCD165, MarUSA) of diameter 165 mm.

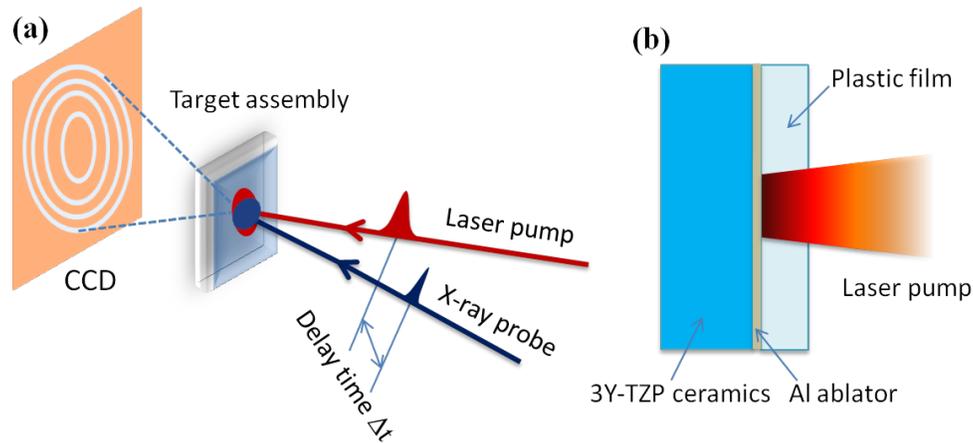

**FIG. 1**. (Color online) (a) Schematic of the laser pump-X-ray probe scheme to measure time-resolved X-ray polycrystalline diffraction of laser-shocked sample. (b) Target assembly used to confine laser-generated plasma and produce pressure pulse in the sample.

Taking laser irradiation conditions into account, we roughly estimated the shock pressure of 5 GPa in 3Y-TZP by using an analytical model of laser-induced plasma in confined geometry.[30, 31] Consequently, the estimated shock wave velocity was lower than 7 km/s.[20, 21] On the other hand, the plasma-confined geometry increased the pressure pulse duration, being longer than the laser pulsewidth. Therefore, the shocked state of 3Y-TZP can maintain much longer than 7 ns, and then is released due to the shock wave reflection at the rear free surface. After multiple-round travelling of waves in sample, finally the mechanical equilibrium is reached. Two normalized intensity profiles integrated from the Debye-Scherrer patterns of laser-shocked 3Y-TZP are shown in Fig. 2(a) and 2(b), respectively corresponding to $\Delta t$=15 ns and 1005 ns. Based on above estimation, we expect that 3Y-TZP was partially at the compressed state in the former case and was completely released in the latter case. In comparison with the correspondent references (normalized intensity profiles of unshocked 3Y-TZP), we observed the appearance and disappearance of new diffraction peaks at $\Delta t$=15 ns and $\Delta t$=1005 ns, respectively. These new peaks are consistent with the monoclinic phase. The subtracted intensity changes give more obvious results, plotted in Fig. 2(c) and 2(d).

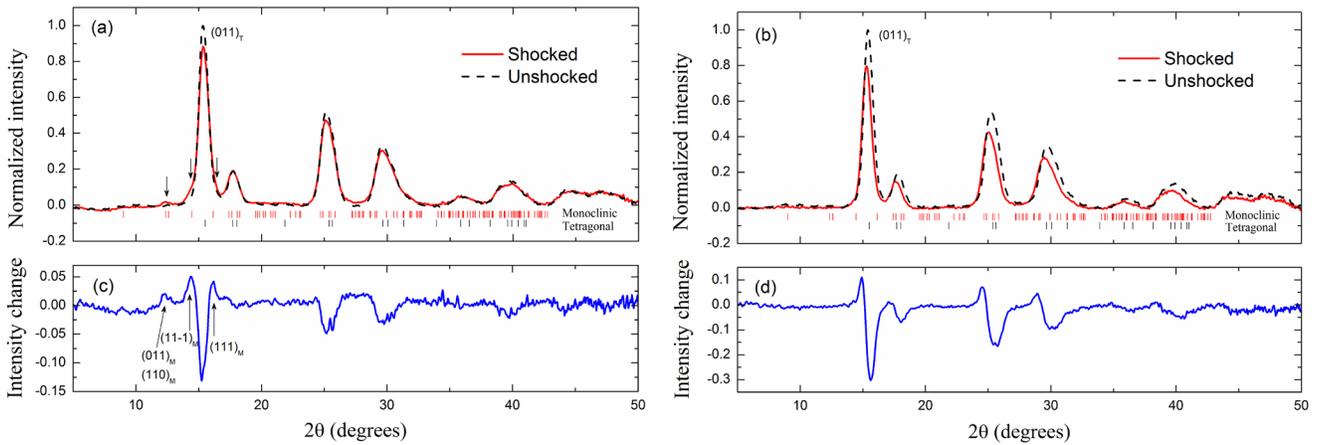

**FIG. 2**. (Color online) Normalized intensity profiles of shocked 3Y-TZP in the cases of $\Delta t$=15 ns (a) and 1005 ns (b), and correspondent intensity changes (c) and (d). Two series of bars at the bottom of (a) and (b) respectively show the calculated diffraction peaks of the monoclinic and tetragonal phases. The subscripts M and T represent the monoclinic and tetragonal phases, respectively.

A series of experiments at different delay time ranging from 5 ns to 1005 ns systematically demonstrated the time evolution of the subtracted intensity profile, shown in Fig. 3. At $\Delta t < 25$ ns, new diffraction peaks corresponding to the monoclinic phase exist, the intensities of which first increase and then decrease, indicating the dynamic process of compressing and releasing. In this time range the sample is at least partially compressed, in which the tetragonal and monoclinic phases coexist. After that, we observed that the diffraction peak (011) of the tetragonal phase gradually shifts to low angle, implying the volume expansion induced by releasing. All facts indicate that the laser-shocked 3Y-TZP ceramics underwent an ultrafast reversible tetragonal-monoclinic transformation, although the shock pressure is much lower than the reported phase transition pressure and the HEL.[17-23] The time scale for this transition is nanosecond or less, which is an important characteristic of martensitic transformation.[12] The crystal structure is reformed by shear stress accompanied by compression, such that the transformation is fast and diffusionless.

Previous shock-recovery investigations suggested that the shock-triggered tetragonal-monoclinic phase transition occurred irreversibly during releasing.[20, 22] The present real-time observation, however, undoubtedly shows that such transformation takes place during compressing, and is reversed during releasing. The similar tetragonal-monoclinic phase transition has also been observed in the static high-pressure experiments of 2Y-TZP by Ohtaka *et al.*, in which the tetragonal phase keeps at the pressure to 14 GPa, and transforms to the monoclinic phase at 14-16 GPa and the orthorhombic II phase over 16 GPa.[17, 18] The observed reversibility, in contrast to the irreversible transformation in the shock-recovery experiments,[21-23] is due to insufficient energy to overcome the energy barrier of transformation. With experimentally available conditions including time-resolution and shock pressure, no intermediate structure from tetragonal to monoclinic is observed during both loading and unloading. Therefore, direct transformation path is reasonably favorable. A possible lattice correspondence is $(100)_M//(110)_T$ and $[001]_M//[001]_T$.[32]

Although the present tetragonal-monoclinic transition has also been observed in several shock-recovery experiments,[21-23] it is introductive to stress that such pressure-induced transition is unexpected in some sense. It is well known that, in pure zirconia, both the monoclinic and tetragonal phases are stable at ambient-pressure, one at room temperature and the other at high temperature (>1500 K). Two phases have different adjacent high-pressure phases: monoclinic to orthorhombic I and tetragonal to orthorhombic II.[15] One may expect that the yttria-stabilized tetragonal phase will directly transform into the orthorhombic II phase. However, the experimental results indicate that the metastable tetragonal phase under compression goes to the monoclinic phase which has the global minimum of the potential energy surface, thus imply that the stabilization effect due to addition of $Y_2O_3$ is negated to some extent by the shear stress under compression.

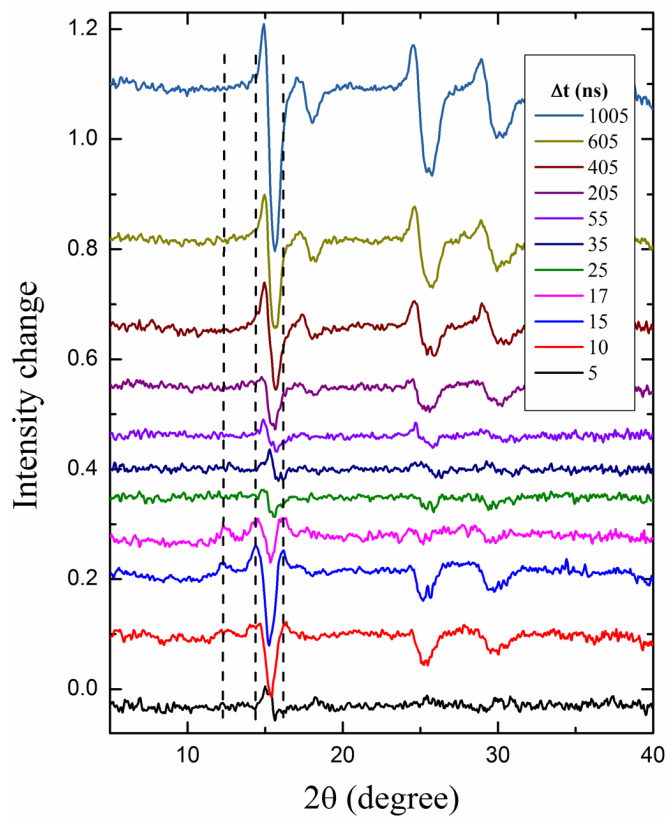

**FIG. 3**. (Color online) Normalized intensity change as a function of the delay time Δt, demonstrating the time evolution of structural dynamics of the laser-shocked 3Y-TZP. The dashed lines indicate the position of observable new diffraction peaks corresponding to the monoclinic phase.

By fitting the diffraction peaks in the angle range of 10 to 19 degree, we extracted the angle shift, $\Delta\theta$, of the main peak (011) of the tetragonal phase, as shown in Fig. 4. Here, to escape from the complexity induced by the diffraction peak overlap of two phases, we started quantitative analysis from the case of $\Delta t=25$ ns, that is, the case in which the transformation has been reversed. Correspondently, the relative interplanar spacing change $\Delta d/d_0$ of the (011) crystal plane is derived according to the differentiation of Bragg's law $\Delta d/d_0 = -\Delta\theta\cot\theta_0$, where $d_0$ and $\theta_0$ are the interplanar spacing and diffraction angle before shock compression, respectively. The existence of the kink is probably due to the reflection and interaction of waves in the sample, which intermittently compresses the sample to some extent till reaching mechanical equilibrium. Due to the shock pressure lower than the HEL, the shocked material is still in the elastic deformation region. Therefore, each grain with different orientation underwent uniaxial compression, leading to the relative volume change $\Delta V/V_0 = \Delta d/d_0$, also shown in Fig. 4. We observed gradually increasing volume expansion along increased delay time. In the elastic regime the residue tensile strain is almost negligible, thus the observed volume change mainly results from thermal expansion. The deduced maximum temperature increment, at $\Delta t=1005$ ns, reaches ~300 K with the volumetric thermal expansion coefficient $\alpha_V \approx 27 \times 10^{-6}$ K$^{-1}$.[33] The estimated temperature increment in 3Y-TZP induced by shock compression at 5GPa, however, is less than 20 K, which cannot cause the observed large volume expansion. The time-dependence of the sample volume, actually, can be understood in term of heat conduction from the laser-induced high-temperature ablation layer to the sample. As pointed out by Fabbro *et al.*,[30] such heat conduction in confined geometry is considerable. To study the temperature effect of shock compression, it is necessary to insulate heat conduction from the ablation layer, which is an important issue in building complete equations of state of materials.

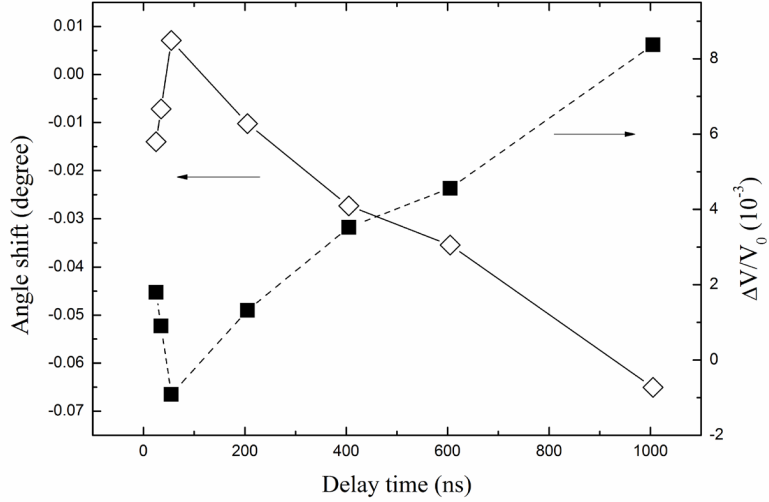

**FIG. 4**. Diffraction peak shift of the crystal plane (011) and deduced relative volume change as a function of the delay time Δt ranging from 25 ns to 1005 ns.

In summary, we demonstrated the ability of time-resolved X-ray diffraction technique to study shock-induced fast phase transition. A reversible martensitic transformation has been observed in the laser-shocked 3Y-TZP ceramics. This work gives new insights into the phase transition dynamics of 3Y-TZP ceramics, such as transformation conditions and pathway. We expect that the progress of such technique will strongly impact the field of shock-wave physics in the future.

## ACKNOWLEDGEMENTS

The authors would like to thank A. Goto, T. Eda, and T. Doki for their help in performing experiments. This work was performed with the approval of the Photon Factory Advisory Committee (Proposal No. 2009G644). J. H. acknowledges financial support from the National Science Foundation of China under grant No. 11064007.